\documentclass[a4paper,10pt]{llncs}
\usepackage{amssymb}
\usepackage{amsmath}
\usepackage{graphicx}
\usepackage[colorlinks=true,linkcolor=blue,anchorcolor=black,citecolor=green,filecolor=red,menucolor=black,urlcolor=blue,breaklinks=true,pdfhighlight=/P,pdfmenubar=true,pdftoolbar=true,pdfpagelabels=true,pdfstartpage=1,pdfstartview=FitV,pdftitle={Titel},pdfsubject={Thema},pdfauthor={Autor},pdfcreator={Ersteller},pdfproducer={Produzent},pdfkeywords={Stichworte}]{hyperref}	
\usepackage[numbers,sort&compress]{natbib}	

\urldef{\mailms}\path|{schultz,fricke}@ifl.tu-dresden.de|
\newcommand{\keywords}[1]{\par\addvspace\baselineskip
\noindent\keywordname\enspace\ignorespaces#1}

\begin{document}
\mainmatter

\title{Stochastic Transition Model for Discrete Agent Movements}

\titlerunning{Stochastic transition model}

\author{Michael Schultz and Hartmut Fricke}

\institute{Institute of Logistics and Aviation, Technische Universit\"at Dresden,\\
Hettner Str. 1-3, D-01062 Dresden, Germany\\
\mailms\\
}

\authorrunning{}

\maketitle

\begin{abstract}

We propose a calibrated two-dimensional cellular automaton model to simulate pedestrian motion behavior. It is a v$_{max}$= 4 (3) model with exclusion statistics and random shuffled dynamics. The underlying regular grid structure results in a direction-dependent behavior, which has in particular not been considered within previous approaches. We efficiently compensate these grid-caused deficiencies on model level.

\keywords{multi-agent, transition matrix, stochastic approach, calibrated model}
\end{abstract}

\section{Introduction}
The different model approaches for microscopic person dynamics are based on the particular discipline analogies, ranging from hydro-dynamic models to artificial intelligence and multi-agent systems \cite{Sch09}. The complex dynamic human behavior is induced by individual decisions, which are classified to be of short-range (operational) and long-range type (strategic/tactical). The self-organization of persons is a further essential characteristic of human behavior. In contrast to the social force model \cite{Hel95,Joh07,Mou10} or the discrete choice model \cite{Bie03,Rob09} the developed motion model \cite{Sch06b,Sch10,Sch10b} is based on a stochastic approach to handle the unpredictable behavior by individual path deviations. The stochastic motion model is an appropriate and fast method for analysis the dynamic pedestrian behavior. However, to derive valid results several simulation runs ($>$100) have to be performed. The focus concentrates on the evaluation of application oriented simulation scenarios instead of the characteristics of individual interactions or specific pedestrian trajectories.

\section{Stochastic Motion Model}
The presented motion model is based on a stochastic approach \cite{Bur01b}, which is comparable to a common cellular automaton. It utilizes a regular grid structure. In contrast to the cellular automaton, the new model is developed on the basis of a fundamental paradigm shift: instead of changing the cell status depending on the status of its surrounding cells (neighbors), the agent is able to move over the regular lattice and to enter those cells, which are not occupied by other agents or obstacles (e.g. walls). To describe the motion behavior of an agent, the motion vector is separated into a desired motion direction and a transversal deviation \cite{Bur01b}. Using the spatially discrete grid structure and defining three transition states (forward $|$ stop $|$ backward or left $|$ on track $|$ right) the normalized transition probability (p) into these states is generally defined by the following equations.

\begin{equation}
\label{eq:define_p}
\begin{split}
	p^+ = \frac{1}{2} \left(\sigma^2 + \mu^2 + \mu \right) \ \ &, \rm{for \ \textit{forward} \ or \ \textit{left}}\\
	p^o = 1 - \left( \sigma^2 + \mu^2 \right) \ \ \ \ \ &, \rm{for \ \textit{stop} \ or \ \textit{on track}}\\
	p^- = \frac{1}{2} \left(\sigma^2 + \mu^2 - \mu \right)  \ \ &, \rm{for \ \textit{backward} \ or \ \textit{right}}\\
\end{split}
\end{equation}

In the case of the desired motion direction, $\mu$ denotes the desired speed and $\sigma^2$ the corresponding variance. If the transversal deviation is concerned, $\mu$ is the average and $\sigma^2$ is the range of the fluctuations. Considering a symmetric transversal deviation ($\mu_{\rm{deviation}}$ = 0) and a connection of desired speed and the corresponding variance (no step backward $p^-$ = 0, so that $\sigma_{\rm{speed}}^2$ = $\mu_{\rm{speed}} (1~-~\mu_{\rm{speed}} )$, the above equations are simplified to the following equations for the desired motion direction (\ref{eq:simple_p1}) and for the transversal motion direction (\ref{eq:simple_p2}).

\begin{equation}
\label{eq:simple_p1}
p^{\rm{forward}} = \mu_{\rm{speed}} \ \  | \ \  p^{\rm{stop}} = 1 - \mu_{\rm{speed}}
\end{equation}
\begin{equation}
\label{eq:simple_p2}
p^{\rm{left, right}} = \frac{1}{2} \sigma_{\rm{deviation}}^2 \ \ \  | \ \  p^{\rm{on~track}} = 1 - \sigma_{\rm{deviation}}^2
\end{equation}

Finally, the motion components are combined to a 3x3 transition matrix ($M_{ij}$) as shown in the following fig. \ref{fig:transitionmatrix}. The emphasized cell (marked gray at the figure) contains the transition probability of moving forward without transversal deviations. In fact, the transition matrix possesses a two-dimensional characteristic, but it only defines an one-dimensional transition considering a transversal deviation (1.5-dimensional).

\begin{figure}
	\centering
	\includegraphics{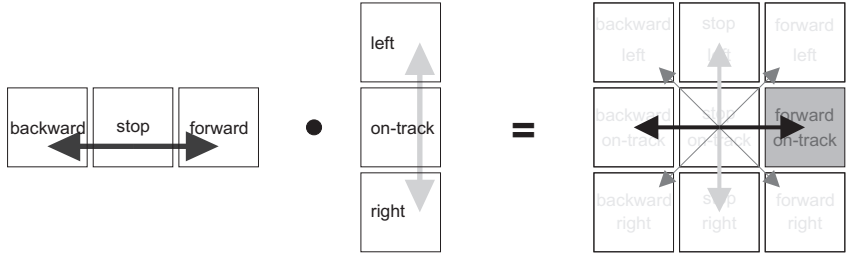}
	\caption{Generation of the transition matrix due to combination of desired motion speed and
transversal motion deviation.}
	\label{fig:transitionmatrix}
\end{figure}

To allow for a three-dimensional agent motion behavior, two independent motion directions are needed. Based on the developed horizontal transition matrix ($\hat{M}$) a diagonal transition matrix ($\tilde{M}$) is derived by re-indexing the horizontal matrix (turning the matrix by $\frac{\pi}{4}$, \cite{Bur01b}). The motion direction ($\alpha$) is integrated into the stochastic model by superpose these matrices with $\lambda$. 

\begin{equation}
\label{eq:matrices}
M = \begin{cases} \left( 1-\lambda \right) \hat{M} + \lambda \tilde{M} \ , & \lambda = \rm{tan} \alpha, 0 \le \alpha < \frac{\pi}{4} \newline
\\ \\
\frac{1}{\sqrt{2}} \left( 1-\lambda \right) \hat{M} + \sqrt{2} \lambda \tilde{M} \ , & \lambda = \rm{tan} \left( \frac{\pi}{4} - \alpha \right), \frac{\pi}{4} \le \alpha \le \frac{\pi}{2} \newline
\end{cases} 
\end{equation}

The rotation of M (4-fold symmetry) allows for determining the entire spectrum of the motion direction. The underlying regular grid structure results in a direction dependent behavior (e.g. entering diagonal cells implies walking a longer way in comparison to horizontally located cells). Therefore the first model modification is to adjust $\lambda$ within the parameters of $\frac{\pi}{4} \le \alpha \le \frac{\pi}{2}$.

\begin{figure}
	\centering
	\includegraphics{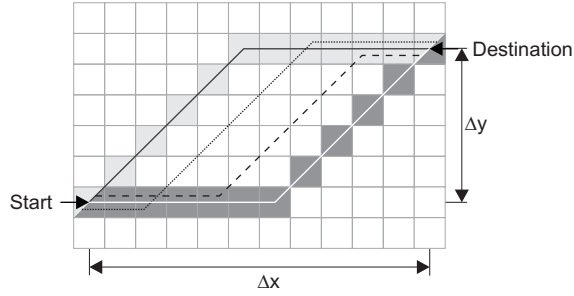}
	\caption{The distance metric $d$ results from the Moore-Neighborhood.}
	\label{fig:model_metric}
\end{figure}

It's obvious, that the stochastic motion model allows for horizontal and diagonal movements (fig. \ref{fig:model_metric}, Moore-Neighborhood).

The shortest distance between two points (start, destination) is given by the metric $d$ (\ref{eq:metric}) which differs from the defined p-norms (Manhattan, Euclidean and Chebychev norm).

\begin{equation}
\label{eq:metric}
d = | \Delta x - \Delta y | + \sqrt{2} \rm{min} \left( \Delta x, \Delta y \right)
\end{equation}

\section{Model Constraints, Improvements and Validation}

Due to the utilization of a regular grid structure, the transition matrix does not fulfill the criteria of independent agent motion behavior. So, the speed and the variance of the agent depend on the agent motion direction. If the agent enters diagonal cells his walking distance is longer (approx. 41\%) in comparison to the use of horizontally located cells. This model constraint is equivalent to a significant higher motion speed depending on the direction of motion. Algorithms to compensate this grid-based speed effect can be found at \cite{Sch06b,Kre07}. A detailed model analysis points out that the expected value of the transition matrix $\mu_{\rm{M}}$  (\ref{eq:trans_matrix}), defined by cell based transition probability $M_{ij}$ and the relative location $\textbf{e}_{ij}$, differs from $\mu_{\rm{speed}}$, which is specified in the motion model. 

\begin{equation}
\label{eq:trans_matrix}
\mu_M = \sum_i{} \sum_j{} \textbf{e}_{ij} M_{ij} \ \ ,  \ \ \ \ \ \  \rm{with} \ \ \textbf{e}_{ij} = \left( \begin{array}{cc} i \\ j \end{array} \right)
\end{equation}

\begin{figure}
	\centering
	\includegraphics{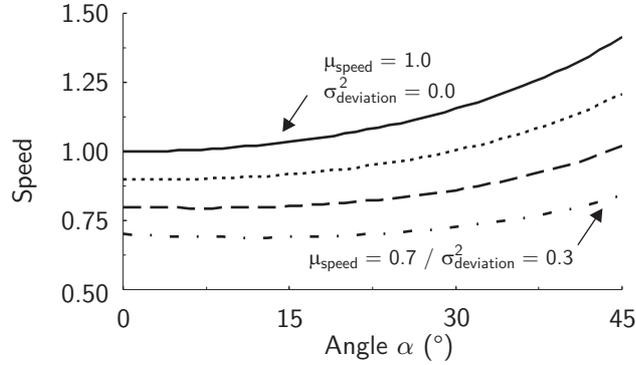}
	\caption{Model deficiency due to motion angle $\alpha$ depending characteristic of speed.}
	\label{fig:error_1}
\end{figure}

Fig. \ref{fig:error_1} points out the correlation of speed and motion angle. With increasing $\alpha$ the corresponding speed ($\mu_{\rm{speed}}$) increases as well, whereas the stepwise change of the transversal deviation ($\sigma_{\rm{deviation}}^2$) mitigates the $\alpha$-depending characteristics.

\begin{figure}
	\centering
	\includegraphics{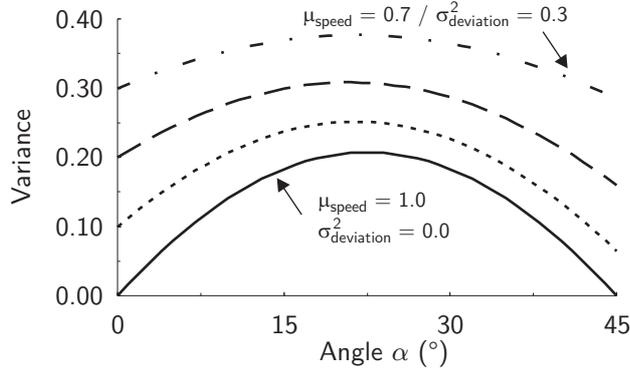}
	\caption{Model deficiency due to motion angle depending characteristic of variance.}
	\label{fig:error_2}
\end{figure}

The stochastic motion model allows for absolute cell transitions ($M_{ij}=1$ with $\sigma_{\rm{deviation}}^2 = 0$) only at motion angle of $\alpha = 0$ and $\alpha = \frac{\pi}{4}$. If the agents choose another angle he always has to choose between two cells at least. Because of the superposition of the horizontal ($\hat{M}$) and diagonal ($\tilde{M}$) transition matrices, a model immanent variance from desired agent motion direction occurs, even if the model defines $\sigma_{\rm{deviation}}^2 = 0$). The characteristic of this overall motion model variance is shown in fig. \ref{fig:error_2}.

The angle depended variance of motion implies to major issues. The model shows an immanent motion deviation, which is not considered in previous equations and the different variances lead to different avoiding behavior. Using the parameter set $\mu_{\rm{speed}}=1$ and $\sigma_{\rm{deviation}}^2 = 0$ as an example, at $\alpha = 0$ no variance is allowed by the model and an agent cannot move if the chosen cell is blocked. Using the same scenario with a different motion angle (e.g. $\alpha = \frac{\pi}{8}$), the agent gets the probability of approx. 20 \% to pass the blocked cell. To ensure homogeneous variance, an appropriate compensation on model level is needed. The expected value $\mu_M$ of the matrix depends on $\mu_{\rm{speed}}$ and $\sigma_{\rm{deviation}}^2$ whereas the parameters are directly coupled. For each parameter set a specific characteristic over the angle has to be calculated. The following fig. \ref{fig:modell_kor} shows these characteristics.
\begin{figure}[b!]
	\centering
	\includegraphics{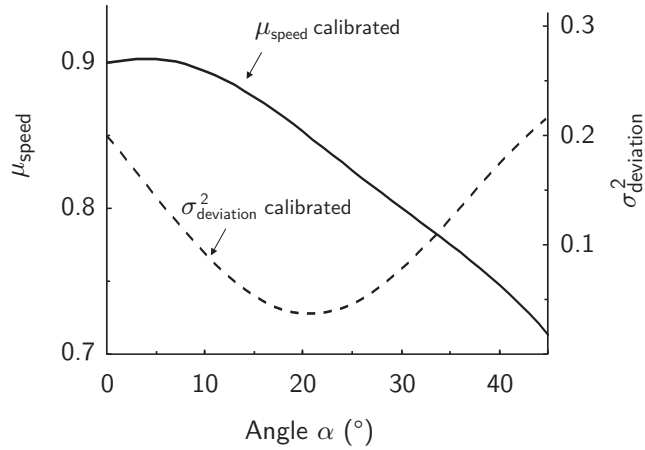}
	\caption{Functional characteristics of the parameter compensation for the parameter set $\mu_{\rm{speed}}=0.9$ and $\sigma_{\rm{deviation}}^2=0.2$.}
	\label{fig:modell_kor}
\end{figure}

\begin{figure}
	\centering
	\includegraphics{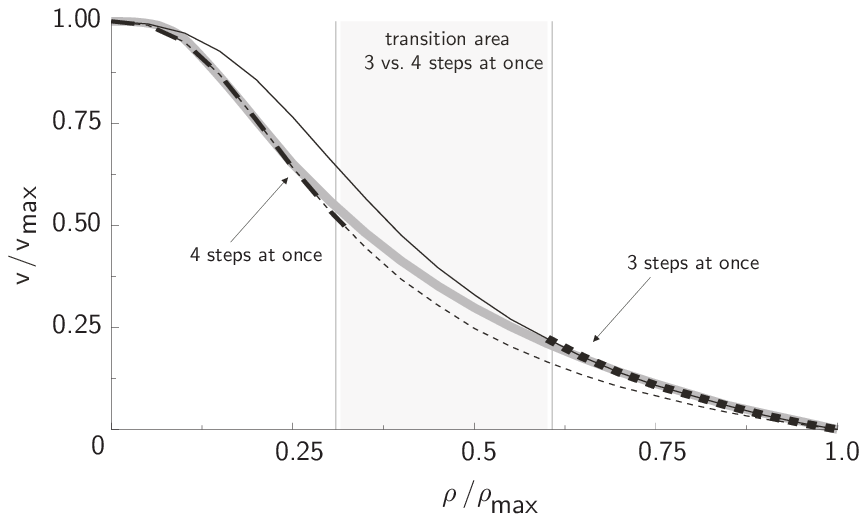}
	\caption{Fundamental diagram of stochastic motion model using different agent operating distances (3-4 steps at once) depending on agent density.}
	\label{fig:fundamental_diagram}
\end{figure}

Further model investigations point out that the model has to be extended to reproduce the representative shape of the fundamental diagram \cite{Wei92,Sey05a}. Using the transition matrix, an agent is able to react to the status of the adjacent cells (empty, occupied). The fundamental diagram indicates an interaction range of about 1.3~m, because the speed of an agent starts to decrease if the density relations $\rho / \rho_{\rm{max}}$ reaches a level of 10~\% (considering an agent with a dimension of 0.4$\times$0.4~m and a maximum density of $\rho_{\rm{max}}=6.25~\rm{Person} / m^2$). If the agent moves three/four steps at once, he will be able to interact with distant agent and the developed model is found to reproduce the characteristic shape of the fundamental diagram (fig. \ref{fig:fundamental_diagram}). Therefore the motion model has to provide the following agent properties:
\begin{itemize}
	\item Always move, no waiting (occupied cells increase transition probability of the other matrix cells).
	\item Move four steps at once and decrease the steps depending on agent density, at least in the case of $\rho$/$\rho_{\rm{max}}>0.6$ the number of steps should be reduced from four to three to fit the shape. 
	\item Agent leaves a trace, at each time step all entered cells will temporarily blocked.
\end{itemize}

The stochastic model meets all criteria for a scientifically reliable motion model. It exhibits the absence of significant model-caused limitations and reproduces all common self-organizing effects (e.g. row formation or oscillation). Besides the operational motion definition by the stochastic transition matrix, strategic/tactical motion components are taken into account as well. The stochastic model allows for the reaction of the agent to objects/agents at immediate vicinity. It additional provides the capability of considering distant constellation of agents (jam) and potentially blocked bottlenecks.

\section{Summary and Outlook}
Model specific parameter corrections ensure that the motion vector is equal to the expected value of the corresponding transition matrix. This issue has in particular not been considered within previous approaches. The calibrated motion model is thus the first approach, which allows for a specific stochastic description of agent movements without model restrictions. 

The passenger related evaluations and simulations of dispatch processes at Dresden Airport exemplarily show that the developed stochastic motion model is able to reproduce the behavior of passengers in an appropriate way \cite{Sch10,Sch10b}. Therefore the developed motion model is implemented in a corresponding application environment which allows for a various application, e.g. investigations regarding to group dynamic behavior or route planning in the airport terminal focused on normal operations and emergency cases.

\bibliographystyle{utphys_quotecomma}
\bibliography{MotionModel_ACRI10}

\providecommand{\href}[2]{#2}\begingroup\raggedright\begin{thebibliography}{10}

\bibitem{Sch09}
A.~Schadschneider, W.~Klingsch, H.~Kluepfel, T.~Kretz, C.~Rogsch, and
  A.~Seyfried, {\em Evacuation Dynamics: Empirical Results, Modeling and
  Applications},
  \href{http://dx.doi.org/10.1007/978-0-387-30440-3_187}{ch.~Evacuation
  Dynamics: Empirical Results, Modeling and Applications, pp.~3142--3176}.
\newblock Springer, 2009.

\bibitem{Hel95}
D.~Helbing and P.~Moln\'ar,
  \href{http://dx.doi.org/10.1103/PhysRevE.51.4282}{``Social force model for
  pedestrian dynamics'',{\em Phys. Rev. E} {\bf 51} (May, 1995)  4282--4286}.

\bibitem{Joh07}
A.~Johansson, D.~Helbing, and P.~Shukla, ``Specification of the social force
  pedestrian model by evolutionary adjustment to video tracking data'', {\em
  Advances in Complex Systems} {\bf 10} (2007) no.~4, 271--288.

\bibitem{Mou10}
M.~Moussaid, N.~Perozo, S.~Garnier, D.~Helbing, and G.~Theraulaz, ``The walking
  behaviour of pedestrian social groups and its impact on crowd dynamics'',
  \href{http://dx.doi.org/10.1371/journal.pone.0010047}{{\em PLoS One} {\bf 5}
  (2010) no.~4, }.

\bibitem{Bie03}
M.~Bierlaire, G.~Antonini, and M.~Weber, ``Behavioral dynamics for
  pedestrians'', in {\em Proceedings of the 10th International Conference on
  Travel Behavior Research}.
\newblock Lucerne, 2003.

\bibitem{Rob09}
T.~Robin, G.~Antonini, M.~Bierlaire, and J.~Cruz, ``Specification, estimation
  and validation of a pedestrian walking behavior model'',
  \href{http://dx.doi.org/DOI: 10.1016/j.trb.2008.06.010}{{\em Transportation
  Research Part B: Methodological} {\bf 43} (2009) no.~1, 36--56}.

\bibitem{Sch06b}
M.~Schultz, S.~Lehmann, and H.~Fricke, {\em Pedestrian and Evacuation Dynamics
  2005}, \href{http://dx.doi.org/10.1007/978-3-540-47064-9_35}{ch.~A discrete
  microscopic model for pedestrian dynamics to manage emergency situations in
  airport terminals, pp.~369 -- 375}.
\newblock Springer, Berlin, 2007.

\bibitem{Sch10}
M.~Schultz, {\em Entwicklung eines individuenbasierten Modells zur Abbildung
  des Bewegungsverhaltens von Passagieren im Flughafenterminal}.
\newblock PhD thesis, Technische Universit\"at Dresden, 2010.

\bibitem{Sch10b}
M.~Schultz, C.~Schulz, and H.~Fricke, {\em Pedestrian and Evacuation Dynamics
  2008}, \href{http://dx.doi.org/10.1007/978-3-642-04504-2_33}{ch.~Passenger
  Dynamics at Airport Terminal Environment, pp.~381 -- 396}.
\newblock Springer, Berlin, 2010.

\bibitem{Bur01b}
C.~Burstedde, K.~Klauck, A.~Schadschneider, and J.~Zittartz, ``Simulation of
  pedestrian dynamics using a 2-dimensional cellular automaton'',
  \href{http://dx.doi.org/10.1016/S0378-4371(01)00141-8}{{\em Physica A} {\bf
  295} (2001)  507--525}.

\bibitem{Kre07}
T.~Kretz and M.~Schreckenberg, {\em Pedestrian and Evacuation Dynamics 2005},
  \href{http://dx.doi.org/10.1007/978-3-540-47064-9_26}{ch.~Moore and more and
  symmetry, pp.~297 -- 308}.
\newblock Springer, Berlin, 2007.

\bibitem{Wei92}
U.~Weidmann, ``{Transporttechnik der Fu\ss g\"anger}'', {\em Schriftenreihe des
  Institut f\"ur Verkehrsplanung, Transporttechnik, Strassen- und Eisenbahnbau}
  {\bf 90} (1992)  .

\bibitem{Sey05a}
A.~Seyfried, B.~Steffen, W.~Klingsch, and M.~Boltes, ``The fundamental diagram
  of pedestrian movement revisited'',
  \href{http://dx.doi.org/10.1088/1742-5468/2005/10/P10002}{{\em J.STAT.MECH.}
  (2005)  P10002}.

\end{thebibliography}\endgroup

\end{document}